\documentclass[aps,prb,twocolumn,a4]{revtex4}

\usepackage{amsmath,amssymb,amsfonts,amsthm}
\usepackage{graphicx}

\begin{document}

\title{Toroidal moments as indicator for magneto-electric coupling:
  the case of BiFeO$_3$ versus FeTiO$_3$}

\date{\today}

\author{Claude Ederer}
\affiliation{School of Physics, Trinity College, Dublin 2, Ireland}  
\email{edererc@tcd.ie}

\begin{abstract}
In this paper we present an analysis of the magnetic toroidal moment
and its relation to the various structural modes in $R3c$-distorted
perovskites with magnetic cations on either the perovskite $A$ or $B$
site. We evaluate the toroidal moment in the limit of localized
magnetic moments and show that the full magnetic symmetry can be taken
into account by considering small induced magnetic moments on the
oxygen sites. Our results give a transparent picture of the possible
coupling between magnetization, electric polarization, and toroidal
moment, thereby highlighting the different roles played by the various
structural distortions in multiferroic BiFeO$_3$ and in the recently
discussed isostructural material FeTiO$_3$, which has been predicted
to exhibit electric field-induced magnetization switching.
\end{abstract}

\maketitle

The concept of magnetic toroidal moments in solids has recently
received increased attention due to its potential relevance in the
context of multiferroic materials and magneto-electric
coupling.\cite{Schmid:2004,VanAken_et_al:2007,Rabe:2007,Ederer/Spaldin:2007,Spaldin/Fiebig/Mostovoy:2008}
A magnetic toroidal moment represents a vector-like electromagnetic
multipole moment which breaks both space and time reversal symmetries
simultaneously. It can be represented by a current flowing through a
solenoid bent into a torus, or alternatively, by a ring-like
arrangement of magnetic dipoles.\cite{Dubovik/Tugushev:1990} The
toroidal moment has been proposed as the primary order parameter for
the low-temperature phase transition from a ferroelectric into a
simultaneously ferroelectric and weakly ferromagnetic,
i.e. multiferroic, phase in boracites.\cite{Sannikov:1997} In
addition, the observation of toroidal domains in LiCoPO$_4$ has
recently been reported.\cite{VanAken_et_al:2007} This suggests that
ferrotoroidicity is a fundamental form of ferroic order, equivalent to
ferromagnetism, ferroelectricity, and
ferroelasticity.\cite{Schmid:2001}

The practical relevance of ferrotoroidic order stems from the fact
that the presence of a magnetic toroidal moment also leads to the
appearance of an antisymmetric magneto-electric
effect.\cite{Gorbatsevich/Kopaev/Tugushev:1983,Sannikov:1997} This is
particularly interesting considering the extensive current research
efforts aimed at finding novel multiferroic materials which exhibit
strong coupling between magnetization and electric
polarization.\cite{Spaldin/Fiebig:2005,Eerenstein/Mathur/Scott:2006,Cheong/Mostovoy:2007,Ramesh/Spaldin:2007}
As suggested in Ref.~\onlinecite{Rabe:2007}, the toroidal moment
concept can offer useful guidance in order to identify possible new
candidate systems and to analyze the specific nature of the
magneto-electric coupling. At the moment, however, it is not fully
clear how this has to be done in practice. It is therefore the purpose
of this work to present an instructive analysis of the toroidal moment
for an important class of multiferroics, and to illustrate how such an
analysis can provide insights into possible coupling between the
various order parameters. Specifically, here we evaluate the toroidal
moment for the case of the $R3c$-distorted perovskite BiFeO$_3$ and
the recently proposed isostructural system FeTiO$_3$ (see
Ref.~\onlinecite{Fennie:2008}).

BiFeO$_3$ is probably the most studied multiferroic to date, whereas
$R3c$ FeTiO$_3$ has only recently been proposed as a material that
exhibits ferroelectrically-induced weak ferromagnetism, and thus
offers the possibility of electric-field controlled magnetization
switching.\cite{Fennie:2008,Ederer/Fennie:2008} First principles
calculations show weak ferromagnetism for both BiFeO$_3$ and $R3c$
FeTiO$_3$.\cite{Ederer/Spaldin:2005,Fennie:2008} The net magnetization
in these systems is due to a slight canting of the mainly
antiferromagnetically ordered Fe spins. This canting is induced by the
Dzyaloshinskii-Moriya
interaction.\cite{Dzyaloshinskii:1957,Moriya:1960} A small
magnetization has indeed been observed experimentally in thin film
samples of BiFeO$_3$,\cite{Eerenstein_et_al:2005,Bea_et_al:2007}
whereas in bulk BiFeO$_3$ this effect is canceled out by the presence
of an additional cycloidal rotation of the antiferromagnetic order
parameter.\cite{Sosnowska/Peterlin-Neumaier/Streichele:1982} As was
shown by both symmetry analysis and explicit first principles
calculations, the weak magnetization is linearly coupled to the
spontaneous electric polarization in FeTiO$_3$, but not in
BiFeO$_3$.\cite{Ederer/Spaldin:2005,Fennie:2008,Ederer/Fennie:2008}
The analysis of the toroidal moment presented in the following
confirms this fact while in addition providing a complementary
perspective.

The toroidal moment $\vec{t}$ of a system of localized magnetic
moments $\vec{m}_i$ at sites $\vec{r}_i$ can be written
as:\cite{Dubovik/Tugushev:1990,Ederer/Spaldin:2007,Spaldin/Fiebig/Mostovoy:2008}
\begin{equation}
\label{toroidal-def}
\vec{t} = \frac{1}{2} \sum_i \vec{r}_i \times \vec{m}_i \quad .
\end{equation}
As described in Ref.~\onlinecite{Ederer/Spaldin:2007}, the presence of
the position vector in Eq.~(\ref{toroidal-def}) together with the
periodic boundary conditions encountered in bulk systems lead to a
multivaluedness of the toroidal moment, in close analogy to the case
of the electric
polarization.\cite{King-Smith/Vanderbilt:1993,Resta:1994} As a result
only differences in the toroidal moment (induced for example by a
structural distortion) are well defined quantities, and the
multivaluedness has to be taken into account when evaluating such
toroidal moment differences. A toroidal state is represented by a
\emph{spontaneous toroidal moment} $\vec{t}_s \ne 0$, where
$\vec{t}_s$ is evaluated as the change in toroidal moment with respect
to a non-toroidal reference configuration. On the other hand, a
non-toroidal state corresponds to a ``centrosymmetric'' ensemble of
toroidal moment values, but does not necessarily imply that the
straightforward evaluation of Eq.~(\ref{toroidal-def}) for one unit
cell leads to $\vec{t}=0$.\cite{Ederer/Spaldin:2007}

\begin{table*}
\caption{Coordinates of all ions $i$ within the rhombohedral unit cell
  of the $R3c$ $AB$O$_3$ structure, $\vec{r}_i = a_1 \vec{a}_1 + a_2
  \vec{a}_2 + a_3 \vec{a}_3$. Without loss of generality we define the
  origin to coincide with the position of the first $A$ cation.}
\label{positions}
\begin{ruledtabular}
\begin{tabular}{c||cc|cc|cccccc}
$i$ & $A$1& $A$2 & $B$1 & $B$2 & O1 & O2 & O3 & O4 & O5 & O6 \\
\hline
$a_1$ & 0 & $\tfrac{1}{2}$ & $(\tfrac{1}{4}+\delta_B)$
&$(\tfrac{3}{4}+\delta_B)$ & $(\tfrac{1}{2}+u)$ & $w$ & $v$ &
$(\tfrac{1}{2}+w)$ & $(\tfrac{1}{2}+v)$ & $u$ \\ 
$a_2$ & 0 & $\tfrac{1}{2}$ & $(\tfrac{1}{4}+\delta_B)$
&$(\tfrac{3}{4}+\delta_B)$ & $v$ & $(\tfrac{1}{2}+u)$ & $w$ &
$(\tfrac{1}{2}+v)$ & $u$ & $(\tfrac{1}{2}+w)$ \\ 
$a_3$ & 0 & $\tfrac{1}{2}$ & $(\tfrac{1}{4}+\delta_B)$
&$(\tfrac{3}{4}+\delta_B)$ & $w$ & $v$ & $(\tfrac{1}{2}+u)$ &
$u$ & $(\tfrac{1}{2}+w)$ & $(\tfrac{1}{2}+v)$ \\ 
\end{tabular}
\end{ruledtabular}
\end{table*}

As already discussed in Ref.~\onlinecite{Ederer/Spaldin:2007}, the
toroidal moment of BiFeO$_3$ evaluated in the limit of localized
magnetic moments vanishes if one takes into account magnetic moments
only on the nominally magnetic Fe sites. This is due to the fact that
the antiferromagnetically ordered Fe cation sublattice in BiFeO$_3$
represents a simple rhombohedral lattice, with inversion centers
located on each cation site, and thus $\vec{t}_s=0$. The same holds
true for $R3c$ FeTiO$_3$. The symmetry-breaking required for a
nonvanishing toroidal moment in these systems is due to the structural
distortions exhibited by the oxygen network surrounding the magnetic
cations. In the following we will therefore assume that small induced
magnetic moments are located on the anion sites in both BiFeO$_3$ and
FeTiO$_3$, and we will evaluate the toroidal moment corresponding to
these induced magnetic moments on the oxygen sites. Note that if the
full magnetization density would be taken into account when evaluating
the toroidal moment, then the full magnetic symmetry of the system
would automatically be included in the calculation. A formalism for
calculating the toroidal moment directly from the quantum mechanical
wavefunction has been suggested
recently.\cite{Batista/Ortiz/Aligia:2008}

\begin{figure}[b]
\includegraphics*[width=0.99\columnwidth]{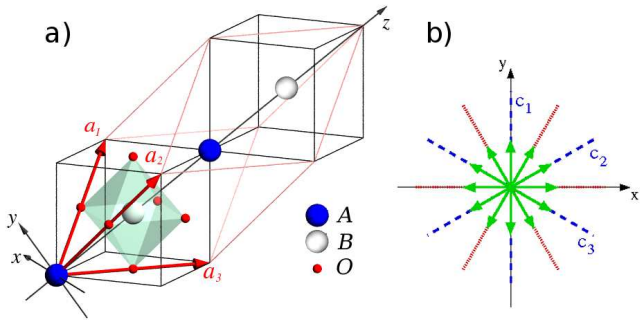}
\caption{a) Unit cell definition used in this work. To better visualize
  the orientation of the coordinate system and rhombohedral basis
  vectors $\vec{a}_i$, the rhombohedral unit cell is inscribed into
  two cubes of the underlying perovskite structure. Only ions within
  one rhombohedral unit cell are shown. The depicted atomic positions
  correspond to the undistorted case. b) Orientation of the three
  glide planes $c_{1,2,3}$, and of the 12 equivalent directions for
  the antiferromagnetic order parameter $\vec{L}$ (arrows) in the $x$-$y$
  plane.}
\label{unit-cell}
\end{figure}

In this work, we are considering perovskite-derived systems with
structural $R3c$ symmetry,\cite{footnote} i.e. the crystal structure
found experimentally for BiFeO$_3$ at ambient
conditions.\cite{Michel_et_al:1969,Kubel/Schmid:1990} $R3c$ FeTiO$_3$
(and MnTiO$_3$) can be synthesized at high pressure, and remains
metastable at ambient conditions, even though the equilibrium crystal
structure in this case is the illmenite structure (space group
$R\bar{3}$).\cite{Ko/Prewitt:1988,Ming_et_al:2006}

We use a rhombohedral setup with lattice vectors defined as:
$\vec{a}_1 = \left( \tfrac{\sqrt{3}}{2}a, \tfrac{1}{2}a, \tfrac{1}{3}c
\right)$, $\vec{a}_2 = \left( - \tfrac{\sqrt{3}}{2}a, \tfrac{1}{2}a,
\tfrac{1}{3}c \right)$, $\vec{a}_3 = \left( 0, -a, \tfrac{1}{3}c
\right)$ (see Fig.~\ref{unit-cell}a). With this choice of coordinate
system, the electric polarization is oriented along the $z$ direction,
whereas the magnetic order parameters will be oriented within the
$x$-$y$ plane.

\begin{figure}
\includegraphics[width=0.5\columnwidth]{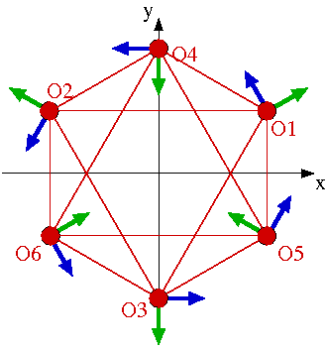}
\caption{(Color online) Displacement vectors corresponding to $\phi$
  (blue/dark grey) and $\epsilon$ (green/light grey) for the six
  oxygen anions in the $R3c$ unit cell, viewed along the $z$
  direction.}
\label{fig:distortions}
\end{figure}

The positions of all ions within the unit cell are listed in
Table~\ref{positions}. The oxygen anions occupy Wyckoff positions $6b$
of the $R3c$ space group. It can easily be seen that
$u=v=w=\delta_B=0$ corresponds to the undistorted ``ideal perovskite''
case (but in our case with $R\bar{3}m$ symmetry, due to the
rhombohedral distortion of the lattice vectors for $c/a \neq
3\sqrt{2}$). In the following it will be convenient to express the
oxygen coordinates in a somewhat different form using
$u=\delta_\text{O} + \frac{2}{3} \epsilon$, $v=\delta_\text{O} -
\frac{1}{3} \epsilon + \frac{1}{\sqrt{3}} \phi$, and
$w=\delta_\text{O} - \frac{1}{3} \epsilon - \frac{1}{\sqrt{3}}
\phi$. In this notation, the oxygen positions are:
\begin{subequations}
\begin{align}
\label{O-pos-begin}
\vec{r}_\text{O1} & = \frac{1}{2} \vec{a}_1 + \left( \begin{array}{c}
  \frac{\sqrt{3}}{2}a \epsilon - \frac{1}{2}a \phi \\ 
  \frac{1}{2}a \epsilon + \frac{\sqrt{3}}{2}a \phi \\
  \delta_\text{O} c \end{array} \right) \quad , \\
\vec{r}_\text{O2} & = \frac{1}{2} \vec{a}_2 + \left( \begin{array}{c} 
  -\frac{\sqrt{3}}{2}a \epsilon - \frac{1}{2}a \phi \\ 
  \frac{1}{2}a \epsilon - \frac{\sqrt{3}}{2}a \phi \\
  \delta_\text{O} c \end{array} \right) \quad , \\
\vec{r}_\text{O3} & = \frac{1}{2} \vec{a}_3 + \left( \begin{array}{c}
  a \phi \\ 
  -a \epsilon \\
  \delta_\text{O} c \end{array} \right) \quad , \\
\vec{r}_\text{O4} & = \frac{1}{2} (\vec{a}_1 + \vec{a}_2) +
\left( \begin{array}{c} 
  -a \phi \\ 
  -a \epsilon \\
  \delta_\text{O} c \end{array} \right) \quad , \\
\vec{r}_\text{O5} & = \frac{1}{2} (\vec{a}_1 + \vec{a}_3) +
\left( \begin{array}{c} 
  -\frac{\sqrt{3}}{2}a \epsilon + \frac{1}{2}a \phi \\ 
  \frac{1}{2}a \epsilon + \frac{\sqrt{3}}{2}a \phi \\
  \delta_\text{O} c \end{array} \right) \quad , \\
\vec{r}_\text{O6} & = \frac{1}{2} (\vec{a}_2 + \vec{a}_3) +
\left( \begin{array}{c} 
  \frac{\sqrt{3}}{2}a \epsilon + \frac{1}{2}a \phi \\ 
  \frac{1}{2}a \epsilon - \frac{\sqrt{3}}{2}a \phi \\
  \delta_\text{O} c \end{array} \right) \quad .
\label{O-pos-end}
\end{align}
\end{subequations}
It can be seen that $\phi$, $\epsilon$, and $\delta_\text{O}$ define
three distinct distortions of the oxygen network: $\delta_\text{O}$
represents the displacement of the oxygen anions along the polar axis
relative to the $A$ site cations, $\phi$ represents the
counter-rotation of the oxygen octahedra around this axis (including a
breathing, i.e. an overall volume change of the octahedra), and
$\epsilon$ represents an additional deformation of the octahedra,
which compresses one side of the octahedron while expanding the
opposing side (see Fig.~\ref{fig:distortions}).

As stated above, for the completely undistorted structure with
$\delta_\text{O}=\delta_B=\epsilon=\phi=0$ the crystallographic
symmetry of the system is $R\bar{3}m$. In this case (and if we for now
do not consider magnetic order) two primitive cells are included in
our unit cell definition. On the other hand, for $\phi \neq 0$ (but
otherwise $\epsilon=\delta_\text{O}=\delta_B=0$) the resulting
symmetry is $R\bar{3}c$, with doubled primitive cell compared to
$\phi=0$, whereas either $\epsilon \neq 0$, $\delta_\text{O} \neq 0$
or $\delta_B \neq 0$ (while all other distortion parameters are zero)
leads to polar $R3m$ symmetry (again with two primitive cells
contained in our unit cell definition if we neglect magnetic order).

Thus, only $\phi \neq 0$ leads to a crystallographic unit cell
doubling compared to the undistorted case (and unrelated to the
magnetic order), whereas both $\delta_\text{O}$ and $\epsilon$ break
space inversion symmetry and therefore represent polar distortions. In
the case of $\delta_\text{O}$ this is intuitively clear, whereas
$\epsilon$ is perhaps not immediately recognized as polar. However,
$\epsilon$ does indeed destroy the inversion symmetry of the system,
and it can easily be verified that $\epsilon$ also creates an electric
polarization $\vec{P} = \sum_i Z^*_{i} \Delta\vec{r}_i$ if the full
Born effective charge tensor (see
Ref.~\onlinecite{Ghosez/Michenaud/Gonze:1998}) is used for
$Z^*_{i}$. (Here, $\Delta\vec{r}_i$ is the change in the position of
oxygen ion $i$ due to $\epsilon \neq 0$.)

It has been correctly pointed out in
Ref.~\onlinecite{deSouza/Moore:2008} that three independent parameters
are required to describe the complete distortion of the oxygen network
within $R3c$ symmetry. In our notation these three parameters are
$\epsilon$, $\omega$, and $\delta_\text{O}$. However, in
Ref.~\onlinecite{deSouza/Moore:2008} the distortion $\epsilon$ was
incorrectly classified as non-polar, which leads to incorrect
conclusions about possible magneto-electric coupling in BiFeO$_3$, as
will become clear in the following. In fact, as pointed out in
Ref.~\onlinecite{Fennie_2:2008}, the two displacement patterns
represented by $\epsilon$ and $\delta_\text{O}$ (and $\delta_B$ as
well) correspond to the same irreducible representation $\Gamma_4^-$
of the original $Pm\bar{3}m$ space group, i.e. they have the same
symmetry properties, and can therefore be viewed as two components of
the same polar distortion.

We now evaluate the toroidal moment resulting from induced magnetic
moments on the oxygen positions listed in
Eqs.~(\ref{O-pos-begin})-(\ref{O-pos-end}). For this we first have to
discuss the symmetry of the magnetically ordered state. First
principles calculations suggest that for both BiFeO$_3$ and $R3c$
FeTiO$_3$ the preferred orientation of the Fe magnetic moments, and
thus the weak magnetization, is perpendicular to
$\vec{P}$.\cite{Ederer/Spaldin:2005,Fennie:2008} This results in 12
energetically equivalent orientations for the antiferromagnetic order
parameter $\vec{L} = \vec{m}_\text{Fe1} - \vec{m}_\text{Fe2}$ within
the $x$-$y$ plane, either parallel or perpendicular to any of the
three $c$-type glide planes of the underlying $R3c$ structure (see
Fig.~\ref{unit-cell}b). (Here, $\vec{m}_\text{Fe1}$ and
$\vec{m}_\text{Fe2}$ are the magnetic moments of the two Fe cations
within the crystallographic unit cell.) The magnetic order thus breaks
the threefold symmetry around the polar $z$ axis and reduces the
crystallographic $R3c$ symmetry to the magnetic symmetry groups $Cc$
or $Cc'$, depending on whether the Fe magnetic moments are parallel or
perpendicular to the remaining $c$-type glide plane. (Here, $C$
indicates a base-centered monoclinic Bravais lattice and $c'$ a glide
plane combined with time reversal.) In the following we will consider
the two representative cases with $\vec{L}$ aligned either along the
$x$ direction ($Cc'$ symmetry) or along the $y$ direction ($Cc$
symmetry).

If the antiferromagnetic order parameter defined by the Fe magnetic
moments is directed along the $x$ direction, i.e. the magnetic
symmetry is $Cc'$, then the $y$-$z$ plane is a $c$-type glide plane
combined with time reversal. This symmetry operation poses the
following restrictions on the magnetic moments $\vec{m}_i$
($i=1,\dots,6$) at the oxygen sites: $m_{6x}=-m_{2x}$,
$m_{6y/z}=m_{2y/z}$, $m_{5x}=-m_{1x}$, $m_{5y/z}=m_{1y/z}$,
$m_{4x}=-m_{3x}$, $m_{4y/z}=m_{3y/z}$. Evaluating
Eq.~(\ref{toroidal-def}) using these relations together with the
oxygen positions (\ref{O-pos-begin})-(\ref{O-pos-end}), and
considering the multivaluedness according to
Ref.~\onlinecite{Ederer/Spaldin:2007}, results in the following $x$
component of the spontaneous toroidal moment:
\begin{equation}
\label{t-cprime}
t_{s,x}^{(Cc')} = \frac{a}{2} \left\{ \sqrt{3} \phi \left( m_{1z}-m_{2z}
\right) + 3 \epsilon \left( m_{1z}+m_{2z} \right) \right\} \quad .
\end{equation}
Here, we have imposed the additional constraint that $m_{3z} = -
m_{1z} - m_{2z}$, to ensure that $\sum_i m_{iz} = 0$. This corresponds
to a decomposition of the full moment configuration into compensated
and uncompensated parts to ensure independence of the calculated
toroidal moment from the choice of origin (see
Ref.~\onlinecite{Ederer/Spaldin:2007}). We note that both $t_y$ and
$t_z$ are vanishing if appropriate multiples of lattice vectors are
added to the atomic positions of the oxygen anions. This means that
the components of the spontaneous toroidal moment along these
directions are zero (see Ref.~\onlinecite{Ederer/Spaldin:2007}).

The corresponding expression for $Cc$ symmetry, i.e. for orientation
of the antiferromagnetic vector $\vec{L}$ along the $y$ direction, is:
\begin{equation}
\label{t-csym}
t_{s,y}^{(Cc)} = \frac{a}{2} \left\{ \phi \left( m_{1z} + m_{2z} -
2m_{3z} \right) - \sqrt{3} \epsilon \left( m_{1z} - m_{2z} \right)
\right\} \quad .
\end{equation}
In this case the symmetry restrictions for the oxygen magnetic moments
are: $m_{6x}=m_{2x}$, $m_{6y/z}=-m_{2y/z}$, $m_{5x}=m_{1x}$,
$m_{5y/z}=-m_{1y/z}$, $m_{4x}=m_{3x}$, $m_{4y/z}=-m_{3y/z}$, and there
is also a nontrivial $z$ component of the toroidal moment. However,
since this component of the toroidal moment does not contribute to the
coupling between $\vec{P}$ and $\vec{M}$ (see below), we only consider
$t_y$.

It can be seen that in general the toroidal moment in $R3c$ BiFeO$_3$
and FeTiO$_3$ is nonzero for both possible magnetic symmetries, and
that it is related to the structural distortions of the oxygen network
($\vec{t}_s=0$ for $\epsilon=\phi=0$). However, the full functional
dependence of $\vec{t}_s$ on $\phi$, $\epsilon$, and $\delta_\text{O}$
can not be seen from Eqs.~(\ref{t-cprime}) and (\ref{t-csym}), since
in general the values of the oxygen magnetic moments will also depend
on these structural parameters (including also $\delta_B$). To gain
further insight into possible magneto-electric coupling we now
consider the case $\epsilon=0$, $\phi\neq 0$, i.e. the paraelectric
reference phase with crystallographic $R\bar{3}c$ symmetry. As has
been pointed out in Refs.~\onlinecite{Fox/Scott:1977},
\onlinecite{Fennie:2008}, and \onlinecite{Ederer/Fennie:2008}, the
presence of a linear magneto-electric effect in the paraelectric
reference phase will lead to a linear coupling between the
\emph{spontaneous} order parameters $\vec{M}_s$ and $\vec{P}_s$. In
contrast, a linear magneto-electric effect in the multiferroic phase
describes the coupling of an additional \emph{induced} component of
polarization or magnetization to the corresponding reciprocal fields
(e.g. $\vec{M}(\vec{E}) = \vec{M}_s + \alpha \vec{E}$). It is
therefore very important to clearly distinguish between the presence
of a linear magneto-electric effect in the para-phase (where
$\vec{M}_s=\vec{P}_s=0$) and in the multiferroic phase ($\vec{M}_s
\neq 0$ and $\vec{P}_s \neq 0$).

We first consider the case $\vec{L} \parallel x$. For BiFeO$_3$ this
results in magnetic $C2'/c'$ symmetry, which leads to the additional
constraint $m_{2z}=m_{1z}$ and thus vanishing toroidal moment
$t_{s,x}^{(C2'/c')}=0$.  For FeTiO$_3$ the resulting symmetry is
$C2/c'$, which requires $m_{2z}=-m_{1z}$ and thus $t_{s,x}^{(C2/c')} =
\sqrt{3} a \phi m_{1z}$.  The difference between BiFeO$_3$ and
FeTiO$_3$ in this case results from the different site symmetries of
the magnetic Fe sites within $R\bar{3}c$ symmetry. In $R\bar{3}c$
BiFeO$_3$ the Fe cation is located on a site with inversion symmetry,
whereas in $R\bar{3}c$ FeTiO$_3$ the inversion centers are located in
between the magnetic cations (see also
Ref.~\onlinecite{Ederer/Fennie:2008}). For the case $\vec{L} \parallel
y$ the resulting symmetry is $C2/c$ (BiFeO$_3$) or $C2'/c$
(FeTiO$_3$). The additional constraints on the oxygen magnetic moments
are $m_{2z}=-m_{1z}$ and $m_{3z}=0$ for $C2/c$ symmetry and
$m_{1z}=m_{2z}$ for $C2'/c$ symmetry, leading to toroidal moment
components $t_{s,y}^{(C2/c)} = 0$ and $t_{s,y}^{(C2'/c)} =
a\phi(m_{1z}-m_{3z})$, respectively. The calculated spontaneous
toroidal moments for the various cases are summarized in
Table~\ref{table-t}.

\begin{table}
\begin{ruledtabular}
\caption{Calculated values of the spontaneous toroidal moment and
  corresponding magnetic symmetry groups for BiFeO$_3$ and FeTiO$_3$
  in the paraelectric $R\bar{3}c$ structure ($\epsilon=0$) for
  different orientation of the antiferromagnetic order parameter.}
\label{table-t}
\begin{tabular}{l|ll|ll}
 & \multicolumn{2}{c|}{BiFeO$_3$} & \multicolumn{2}{c}{FeTiO$_3$} \\
\hline
$\vec{L} \parallel \hat{x}$ & $C2'/c'$ & $\vec{t}_s = 0$ & $C2/c'$ &
$t_{s,x} = \sqrt{3} a \phi m_{1z}$ \\
$\vec{L} \parallel \hat{y}$ & $C2/c$ & $\vec{t}_s = 0$ & $C2'/c$ &
$t_{s,y} = a \phi (m_{1z}-m_{3z})$ \\
\end{tabular}
\end{ruledtabular}
\end{table}

It can be seen that in the paraelectric $R\bar{3}c$ phase (i.e. for
$\epsilon=0$) only FeTiO$_3$, but not BiFeO$_3$, has a nonvanishing
toroidal moment, and that the toroidal moment in paraelectric
FeTiO$_3$ is related to the counter-rotations of the oxygen octahedra
represented by $\phi$. The presence of this toroidal moment causes a
linear magneto-electric effect $\vec{M}=\alpha \vec{E}$ with $\alpha
\propto \vec{t}$ via the free-energy invariant $E_{TPM} \propto
\vec{t} \cdot \left( \vec{P} \times \vec{M} \right)$ (see
e.g. Ref.~\onlinecite{Ederer/Spaldin:2007}).  This means that once the
polarization in $R\bar{3}c$ FeTiO$_3$ becomes nonzero (which of course
reduces the crystallographic symmetry to $R3c$), it will induce a weak
magnetization via the linear magneto-electric effect, consistent with
the design criteria outlined in Ref.~\onlinecite{Fennie:2008}. Such
``ferroelectrically-induced ferromagnetism'' via the linear
magneto-electric effect has been originally suggested in
Ref.~\onlinecite{Fox/Scott:1977}.

On the other hand, paraelectric $R\bar{3}c$ BiFeO$_3$ is non-toroidal
and does not exhibit a linear magneto-electric effect. Therefore, the
weak magnetization in BiFeO$_3$ is not ferroelectrically-induced and
there is no linear coupling between $\vec{M}_s$ and $\vec{P}_s$ in the
multiferroic phase. This is also consistent with first principles
calculations, where for BiFeO$_3$ weak ferromagnetism occurs in both
the ferroelectric $R3c$ and the paraelectric $R\bar{3}c$ structures,
whereas for FeTiO$_3$ it occurs only in the ferroelectric $R3c$
structure.\cite{Ederer/Spaldin:2005,Fennie:2008}

Note that in the multiferroic $R3c$ phase both FeTiO$_3$ and BiFeO$_3$
exhibit a toroidal moment (according to Eqs. (\ref{t-cprime}) and
(\ref{t-csym})) and thus a linear magneto-electric effect. This means
that an external electric field will induce changes in both
polarization and magnetization, linear in the external field, but only
in FeTiO$_3$ the corresponding spontaneous order parameters
$\vec{M}_s$ and $\vec{P}_s$ are linearly coupled. Such linear coupling
between $\vec{P}_s$ and $\vec{M}_s$ is required to achieve full
electric-field control of the weak magnetization. As outlined in
Refs.~\onlinecite{Fennie:2008} and \onlinecite{Ederer/Fennie:2008} a
reversal of $\vec{P}_s$ in FeTiO$_3$ induced by an external electric
field will result in a corresponding reversal of $\vec{M}_s$ provided
the antiferromagnetic order parameter (or equivalently the toroidal
moment) is fixed by a large enough magnetic anisotropy. On the other
hand, such electric field controlled switching of the weak
magnetization is not expected to occur in BiFeO$_3$.

The evaluation of the toroidal moment presented above allows to
clearly identify which structural modes, in combination with the
antiferromagnetic order, lead to the appearance of a toroidal moment
and a linear magneto-electric effect. In contrast to the
antiferromagnetic order parameter, which generally depends only on the
orientation of the individual magnetic moments, it follows from
Eq.~(\ref{toroidal-def}) that the toroidal moment contains information
about where the magnetic moments are located as well as on how they
are oriented. Furthermore, the toroidal moment is a macroscopic
multipole moment that is related to the (magnetic) point group
symmetry, whereas a proper symmetry analysis of antiferromagnetic
order requires a treatment based on the full space group symmetry. In
particular, antiferromagnetic order is not connected to any particular
macroscopic symmetry breaking, i.e. all 90 magnetic point groups are
compatible with the existence of antiferromagnetic order. On a
microscopic space group level, antiferromagnetic order of course
always breaks time reversal symmetry. However, for systems where the
magnetic unit cell is a multiple of the crystallographic unit cell, a
primitive translation of the original nonmagnetic lattice can be
combined with time reversal, and as a result the corresponding
magnetic point group still contains time reversal as a symmetry
element.\cite{Birss:Book} In contrast, a toroidal moment always breaks
space and time reversal symmetries on the macroscopic level, i.e. the
corresponding magnetic point group does not contain neither space
inversion $\bar{1}$ nor time reversal $1'$ (whereas the combined
operation $\bar{1}'$ can still be a symmetry element). Since all
macroscopic properties of a particular crystal are determined by its
point group rather than space group symmetry,\cite{Birss:Book} the
toroidal moment appears to be a more appropriate quantity to classify
macroscopic symmetry properties compared to the antiferromagnetic
order parameter.  In particular, the toroidal moment is ideally suited
to discuss magneto-structural or magneto-electric coupling. For a
given structural and magnetic configuration the toroidal moment can be
evaluated straightforwardly, applying the procedure outlined in
Ref.~\onlinecite{Ederer/Spaldin:2007}. Using the relation between the
toroidal moment and the magneto-electric tensor $\alpha$, this allows
to correctly identify which quantities determine the magneto-electric
properties of the system. The same can of course also be achieved by a
group theoretical analysis of the symmetry properties of the various
structural modes and of the antiferromagnetic order parameter. The
straightforward evaluation of the toroidal moment should therefore be
considered as an alternative (or complementary) way to discuss
magneto-electric symmetry that does not necessarily require the
application of group theoretical concepts.

Finally, we point out that the toroidal moment is only related to the
antisymmetric part of the linear magneto-electric tensor $\alpha$,
whereas the symmetric part of $\alpha$ is connected to other
electromagnetic multipole moments (see
Ref.~\onlinecite{Spaldin/Fiebig/Mostovoy:2008}). However, for the
present case where weak ferromagnetism is caused by the
Dzyaloshinskii-Moriya interaction, the antisymmetric component related
to the toroidal moment is indeed the crucial part of $\alpha$.

In summary, we have shown that by evaluating the toroidal moment in
the limit of localized magnetic moments, a clear picture of the
different roles played by the various structural distortions for the
magneto-electric properties in BiFeO$_3$ and $R3c$ FeTiO$_3$ can be
achieved. The toroidal moment can be used to characterize the
magneto-electric properties in antiferromagnetic systems. Its
usefulness stems from the fact that is depends on both position and
orientation of the magnetic moments and from its well-defined
macroscopic symmetry properties, which allow to use point groups
instead of space groups, in contrast to a discussion based on the
antiferromagnetic order parameter.

\begin{acknowledgments}
This work was supported by Science Foundation Ireland under
Ref.~SFI-07/YI2/I1051.
\end{acknowledgments}

\bibliography{../../literature.bib,footnotes.bib}

\begin{thebibliography}{34}
\bibitem[{foo()}]{footnote}
\bibinfo{note}{Depending on the amount of distortion, the corresponding
  structure is sometimes referred to as the ``ferroelectric LiNbO$_3$
  structure''. This distinction between distorted perovskite and LiNbO$_3$
  structure is purely quantitative and thus not relevant in the present
  context.}
\end{thebibliography}

\end{document}